\documentclass[granma]{svjour}
\usepackage{epsfig}
\usepackage{amsmath}
\usepackage{amssymb}

\newcommand{\mayapr}{\raisebox{-.4ex}{$\enskip\stackrel{\raisebox{-.8ex}{$\scriptstyle >$}}{\scriptstyle\sim}\enskip$}}
\newcommand{\menapr}{\raisebox{-.4ex}{$\enskip\stackrel{\raisebox{-.8ex}{$\scriptstyle <$}}{\scriptstyle\sim}\enskip$}}

\begin{document}
\title{Dynamics of deviations from the Gaussian state in a freely
  cooling homogeneous system of smooth inelastic particles}
\subtitle{}
\author{M. Huthmann,   J. A. G. Orza and R. Brito%
\thanks{The authors thank to A.~Zippelius, T.~Aspelmeier, 
P.~M\"uller, and A.~Santos for useful discussions. M.~H. acknowledges financial
support by the DFG through  SFB 345 (Germany), and 
J.~A.~G.~O. and R.~B. from DGES number PB97-0076 (Spain).}%
}                     
%
%
\institute{Martin Huthmann, \\
 Institut f\"ur Theoretische Physik, Universit\"at G\"ottingen, 
Bunsenstr. 9, 37073 G\"ottingen, GERMANY\\
R. Brito, J. A. G. Orza\\Dpto. F\'\i sica Aplicada I,
Facultad de Ciencias F\'\i sicas, Universidad Complutense, 28040
Madrid, SPAIN}
\date{Received: / Revised version: }
%
\maketitle
\begin{abstract}
  The time dependence of deviations from the Gaussian state in a
  freely cooling homogeneous system of smooth inelastically colliding
  spheres is investigated by kinetic theory.  We determine the full
  time dependence of the coefficients of an expansion around the
  Gaussian state in Generalized Laguerre polynomials. Approximating
  this system of equations to sixth order, we find that
  the asymptotic state, where the mean energy $T$ follows Haff's law
  with time {\em independent} cooling rate, is reached within a 
  few collisions per particle. 
  Two-dimensional molecular dynamics simulations confirm our results and show
  {\em exponential} behavior in the high-energy tails.
\end{abstract}

\noindent {\bf Keywords:} Inelastic particles, velocity 
distribution, non-Gaussian behavior, high energy tails.

\section{Introduction}
Freely cooling systems of smooth inelastically colliding spheres or
discs (in the following denoted by inelastic hard spheres systems
(IHS)) have been investigated by means of kinetic theory and computer
simulations by several groups (see e.g. 
\cite{goldhirsch,goldhirsch2,deltour,esipoe,mcnamara,noijeprl,Orza}). 
Most of the studies focus on latest
times where interesting phenomena like formation of vortex patterns
\cite{mcnamara,noijeprl}
and clustering \cite{goldhirsch,deltour,Cahn} can be observed.  
For short times or not too high inelasticities, however,  
the system remains homogeneous with a decreasing temperature, 
or equivalently, a decreasing average velocity.
It is the so called Homogeneous Cooling State (HCS) 
the regime that will be studied in this paper. 
The HCS admits a {\em scaling solution}, i.e. if one scales all velocities
with the average velocity and assumes that the shape of the scaled
velocity distribution function remains constant in time, the entire
time dependence is given by the time dependence of the average
velocity only. This scaling solution is the starting point for a
hydrodynamic analysis. Although many of the existing theories use a
Gaussian velocity distribution function (which may be valid for small
inelasticity) in general the shape is not Gaussian. First evidence, at
very late evolution stages, was
obtained by Goldhirsch {\em et al.} \cite{goldhirsch2} by measuring
the fourth moment of the velocity distribution function.  Later
Goldshtein and Shapiro \cite{goldshtein} proposed a solution based on
an expansion in Sonine polynomials. Van Noije and Ernst \cite{noije}
correctly calculated the first term to linear order and Brilliantov
and P\"oschel  \cite{poeschel} included nonlinear corrections.  Numerical
solutions of the Boltzmann equation were calculated using Direct
Simulation Monte Carlo method (DSMC) \cite{brey}.
Finally, extensions to  viscoelastic particles have been recently presented in 
\cite{brilliantov}. 

Further confirmation of non Gaussian behavior is
given by Esipov and P\"oschel \cite{esipoe} by 
studying the high-velocity tails of the velocity distribution
function. They find that the tails are  of an exponential type instead of
a Gaussian, results confirmed by  DSMC by Brey {\em et al.} 
\cite{brey2} and also in experiments performed by Losert and
coworkers \cite{losert}.  In fact, an asymptotic exponential
tail of the form $\exp(-v^\beta)$ with $\beta<2$  seems to be a more 
fundamental behavior, as it is also 
present in driven or vibrated granular materials \cite{noije,losert}.
Recently, a detailed study by DSMC in driven systems with three different 
types of forcing has been presented \cite{Andres}. Only for the
so-called Non-Gaussian thermostat (where there is a balance
between energy input and dissipation) $\beta=2$, while for the other
forcings, $\beta=3/2$ and 1. 
Surprisingly, no molecular dynamics results have been presented so far
for calculating moments and high energy tails for the freely evolving case. 

Our starting point is similar to \cite{goldshtein}, i.e. 
expand the scaled velocity distribution function in a series of 
Generalized or Associated Laguerre polynomials 
around the Gaussian distribution
with  coefficients denoted with $a_l$.  However,
we assume that the coefficients are time dependent \cite{poeschel}.
With these ideas we try to achieve two goals. Firstly, investigate the 
influence of
higher coefficients $a_3,\ldots,a_6$, in the expansion of the 
distribution function.  Secondly study their time evolution. 
We find that for not too high inelasticities the above expansion 
seems to be convergent. Furthermore, the 
cooling proceeds in two stages: (1) A fast decay (in the order of 
few collisions per particle) of all
coefficients $a_l$ to their asymptotically constant values. (2) An
algebraically slow decay of the kinetic energy $T$ determined by
Haff's law $\frac{d}{dt} T = C T^{-3/2}$ and time independent
coefficient $C$ 
depending on the asymptotic values of $a_l$. 
Two dimensional event driven molecular dynamics simulations are performed in
order to test the theory with good agreement for moderate
inelasticity. 
For higher inelasticity the perturbations expansions seems to fail and
in the simulations we were able to observe a transition to an
exponential high-energy tail.

The paper is organized as follows: In Sec.~2 we propose the expansion
of the velocity distribution in Laguerre polynomials with time
dependent coefficients.  In Sec.~3 we determine formally the {\em
  full} time dependence of the HCS expressed by the time evolution of
the coefficients of the expansion. We obtain an infinitely large
system of ordinary differential equations, which can only be
investigated approximately.  This is done in Sec.~4, where a
truncation scheme is proposed and analyzed to different orders. In
Sec.~\ref{simr} we compare the analytical theory to results from
event-driven simulations and the validity of the perturbation
expansion is discussed. Results for the exponential high-energy tail
are presented here.  We summarize the results in Sec.~\ref{conc}.

\section{The system under consideration}
\label{tsuc}
We consider a system of $N$ smooth, inelastically colliding spheres
with diameter $\sigma$ confined to a $d$-dimensional volume $V$, so that
the homogeneous density is given by $n:=\frac{N}{V}$.  The positions
of each sphere are denoted by $\vec{r}_i$ and each particle has a
velocity $\vec{v}_i$.  The particles interact via a hard-core potential
and in each collision (i.e. if $r_{ij} :=
|\vec{r}_{ij}|:=|\vec{r}_i-\vec{r}_j|= \sigma$) the velocities are
instantaneously changed by the following collision rules determined by
a constant coefficient ${e_n}\in[0,1]$ of restitution
\begin{align} \label{coll}
\vec{v}_i' &= \vec{v}_i - \frac{1 + {e_n}}{2}
(\vec{v}_{ij}\cdot \hat{\vec{r}}_{ij})\hat{\vec{r}}_{ij}~,\nonumber  \\ 
\vec{v}_j' &= \vec{v}_j +\frac{1
+ {e_n}}{2} (\vec{v}_{ij}\cdot \hat{\vec{r}}_{ij})\hat{\vec{r}}_{ij}~, 
\end{align}
where $ \vec{v}_{ij}:=\vec{v}_{i}-\vec{v}_{j} $ and
$\hat{\vec{r}}_{ij}=\vec{r}_{ij}/r_{ij}$. Velocities {\em after} collision
are  primed  quantities given by  velocities {\em before} collision (unprimed 
quantities).

The IHS is described statistically 
by the single particle distribution function
$\rho(\vec{r},\vec{v},t)d{\vec{r}} d{\vec{v}} $, the (average) number of particles 
at positions between ${\vec{r}}$ and ${\vec{r}}+d{\vec{r}}$ and 
with velocities between ${\vec{v}}$ and ${\vec{v}}+d{\vec{v}}$
at time $t$. As proposed in \cite{goldshtein}, 
for a homogeneous inelastic system the distribution function
can be expressed by a scaling function as: 
\begin{equation} 
\rho(\vec{v},t) := n \frac{1}{(v_0(t)\sqrt{\pi})^d}
  \tilde{\rho}(\vec{c},t) ,
\end{equation}
where $\vec{c} = \vec{v}/v_0(t)$ and $v_0(t)$ is the thermal
velocity defined as the square root of the second moment
of the distribution function:
\begin{equation}
  \label{eq:ass2}
  \int d\vec{v}\,  \rho(\vec{v},t) \vec{v}^2 = n \frac{d}{2} v_0^2(t)~.
\end{equation} 
The temperature is then defined as $T(t):=\frac{m}{2} v_0^2(t)$. 
For elastic systems the distribution function is Gaussian
and it is expected that it will remain close to a
Gaussian for small inelasticity.
Therefore, we expand the scaled distribution function
in a series of Generalized or Associated Laguerre polynomials 
\cite{oberhettinger} around the Gaussian 
distribution function. The expansion is carried out in the 
scaled velocity variable $\vec{c}$ and with
{\em time dependent coefficients} $a_l(t)$ \cite{poeschel}. 
The general ansatz for the
single particle distributions function for the homogeneous cooling
then reads
\begin{equation}\label{rhotilde}
  \tilde{\rho}(\vec{c},t) :=
  \exp\left(-\vec{c}^2\right) \sum_{l=0}^\infty
  a_l(t) L_l^{\alpha} \left(\vec{c}^2 \right)~,
\end{equation}
where $\alpha=d/2-1$ in $d$ dimensions.         
In the context of kinetic theory 
Laguerre polynomials are called Sonine polynomials
\cite{chapcowl}.

The normalization condition for $\rho$,
$\int d\vec{v} \rho=n$, leads to $a_0=1$.
We express $\vec{v}^2$  by the first and
second Laguerre polynomial 
\begin{equation}
  \vec{v}^2 = - L_1^\alpha(\vec{v}^2)+(\alpha+1)
  L_0^\alpha(\vec{v}^2)~,
\end{equation}
and  using the  orthogonality relations for the Laguerre polynomials
we find 
\begin{equation}
  \int d \vec{v}\, \rho(\vec{v},t) \vec{v}^2 = n v_0^2
  \left(\frac{d}{2}-\binom{1+\alpha}{1} a_1\right)~,
\end{equation}
which implies together with Eq.~(\ref{eq:ass2}) that $a_1=0$ for all
times \cite{goldshtein,noije,poeschel}.
We denoted the binomial coefficients by $\binom{a}{b}$.

Finally, as Laguerre polynomials are orthogonal, the coefficients
$a_l$ are given by
 \begin{equation}\label{eq:aih}
n a_l (t) =\frac{1}{ \binom{l+\alpha}{l}} \int d \vec{v}\, \rho(\vec{v},t)
 L_l^\alpha \left(\left(\frac{\vec{v}}{v_0(t)}\right)^2\right)~.
\end{equation}

\section{The Homogeneous Cooling State}
\label{thcs}
\subsection*{The Boltzmann Equation}
We assume that dynamics of the one particle distribution function
$\rho$ is given by the Enskog Boltzmann equation, which can be written
in $d$ dimensions without external forces as
\begin{equation}\label{eq:bgl}
 \partial_t \rho(\vec{r},\vec{v}_1,t) +
 (\vec{v}_1\cdot\vec{\nabla}_{\vec{r}})\rho(\vec{r},\vec{v}_1,t) =
 J[\rho,\rho]~,
 \end{equation}
with collision integral
 \begin{multline}\label{eq:bgl1}
 J[\rho,\rho] = \sigma^{d-1} \chi \int d\vec{v}_2  \int d
\hat{\vec{\sigma}}
\Theta(\vec{v}_{12}\cdot 
\hat{\vec{\sigma}})(\vec{v}_{12}\cdot\hat{\vec{\sigma}})
\\ \left(\frac{b^{-1}}{{e_n}^2}-1\right) \left( \rho(\vec{r},\vec{v}_1,t)
\rho(\vec{r},\vec{v}_2,t) \right) .
\end{multline}
$\hat{\vec{\sigma}}$ is the unit vector pointing from particle 2 to particle 1,
$\chi$ the pair correlation function at contact, and $b^{-1}$
describes `restituting collisions' by changing velocities in $\rho$,
i.e.  $b^{-1} \rho(\vec{r},\vec{v}'',t) =\rho(\vec{r},\vec{v},t) $ in
a way that $\vec{v}''$ are the velocities before collision leading to
$\vec{v}$ after collision.  The operator $b$ describes `direct
collisions' given in Eqs.~(\ref{coll}). The inverse operator of $b$,
 i.e.  $b^{-1}$, is simply given by substituting ${e_n}$ by $1/{e_n}$ in
Eqs.~(\ref{coll}).

By multiplying the Boltzmann equation Eq.~(\ref{eq:bgl}) with some function
$\psi(\vec{v}_1)$ and integrating over $\vec{v}_1$  one gets 
\begin{multline}\label{eq:bgl2}
 \partial_t \int d\vec{v}_1 \psi(\vec{v}_1)
 \rho(\vec{r},\vec{v}_1,t) +\vec{\nabla}\cdot\int d\vec{v}_1
\vec{v}_1\psi(\vec{v}_1)
\rho(\vec{r},\vec{v}_1,t)  = \\
\int  d\vec{v}_1  \psi(\vec{v}_1) J[\rho,\rho] ~,
 \end{multline}
which can be rewritten in the form of  a balance  equation 
\begin{equation}\label{eq:bgl3}
 \partial_t \overline{\psi} + \vec{\nabla} \cdot\vec{j}_\psi = 
  \partial_t^{\rm coll} \overline{\psi}~,
\end{equation}
describing that the time 
change of an averaged quantity  $\overline{\psi}$ is due to flux
$\vec{j}_\psi $ or due to change through collisions. 
The right hand side of Eq.~(\ref{eq:bgl2}) can be written as \cite{chapcowl}
\begin{multline}\label{eq:bgl4}
  \int  d\vec{v}_1  \psi(\vec{v}_1)
J[f,f] = 
 \sigma^{d-1} \chi \int d\vec{v}_2   d\vec{v}_1  \int d
\hat{\vec{\sigma}} \\
\Theta(\vec{v}_{12}\cdot 
\hat{\vec{\sigma}})(\vec{v}_{12}\cdot\hat{\vec{\sigma}})
\rho(\vec{v}_1)\rho(\vec{v}_2)
\Delta \psi ~,
\end{multline}
and $\Delta \psi$ is the change of $\psi$ in a direct collision for
both particles $\Delta \psi= \frac{1}{2}
(\psi(\vec{v}_1')+\psi(\vec{v}_2')-\psi(\vec{v}_1)-\psi(\vec{v}_2))$.

\subsection*{Dynamics of Moments}
Using Eqs. (\ref{eq:ass2}) and (\ref{eq:bgl4}) the time dependence of
$T(t)= \frac{m}{2} v_0^2 $ in the homogeneous case is given by
\begin{equation} \label{eq:mom1}
  \frac{d}{dt} T =  \frac{d}{dt} \frac{m}{2} v_0^2 =  -2 \gamma \omega_0 T ~,
\end{equation}
where $\gamma$ is defined as 
\begin{multline}
\gamma :=  - \frac{\sqrt{2\pi}}{d S_d}\frac{1}{\pi^d}
  \int d \vec{c}_1 d \vec{c}_2 d
  \hat{\vec{\sigma}} \Theta(\vec{c}_{12}\cdot
  \hat{\vec{\sigma}})(\vec{c}_{12}\cdot\hat{\vec{\sigma}}) \times \\
   \tilde{\rho}(\vec{c}_1) \tilde{\rho}(\vec{c}_2) (b-1)
  \frac{1}{2} (\vec{c}_1^2+\vec{c}_2^2)~,
\label{eq:gg}
\end{multline}
and $\tilde{\rho}$ as in Eq.~(\ref{rhotilde}).  $S_d$ is the surface
of a unit sphere in $d$ dimensions and $\omega_0$ the Enskog collision
frequency for a classical gas of hard spheres with temperature $T$,
given by:
\begin{equation}
\quad S_d = \frac{2 \pi^{d/2}}{  \Gamma(d/2)} \quad {\mbox{and}}\quad
\omega_0 = \frac{S_d}{\sqrt{2 \pi}} \chi n (2\sigma)^{d-1} v_0 ~.
\end{equation}
If the velocity distribution function $\tilde\rho$ in 
Eq.~(\ref{eq:gg}) is a Maxwellian, $\gamma$ takes the value of 
$\gamma_0:= (1-e_n^2)/(2d)$. 
This is the Gaussian  value of 
the energy decay rate obtained by  Haff \cite{haff}. 
The fact that  this distribution function is not a 
Gaussian modifies the cooling rate, as it will be calculated later on. 
 
In order to obtain the time evolution of $a_l$, we take 
the time derivative of Eq.~(\ref{eq:aih}), and it has  to be considered the time
dependence of $\rho(\vec{v},t)$ as well as the time dependence of $
L_l^\alpha ((\frac{\vec{v}}{v_0(t)})^2)$ via $v_0(t)$. The time
dependence of $\rho(\vec{v},t)$ is given by Boltzmann equation, and
the time dependence of $v_0(t)$ is given by equation (\ref{eq:mom1}).
After a straight forward calculation using differential formulas for
the Laguerre polynomials, we get
\begin{equation} \label{tima}
\frac{d}{dt} a_l  = \omega_0 \gamma_l + l 2 \gamma  \omega_0 (a_l -a_{l-1})~,
\end{equation}
and 
\begin{multline}
\gamma_l = \frac{\sqrt{2\pi}}{S_d}\frac{1}{\binom{l+\alpha}{l}} \frac{1}{\pi^d} \int d
 \vec{c}_1 d \vec{c}_2 d \hat{\vec{\sigma}} \Theta(\vec{c}_{12}\cdot 
\hat{\vec{\sigma}})(\vec{c}_{12}\cdot\hat{\vec{\sigma}}) \times \\
 \tilde{\rho}(\vec{c}_1)\tilde{\rho}(\vec{c}_2) (b-1)\frac{1}{2}
 (L_l^{\alpha}(\vec{c}_1^2) + L_l^{\alpha}(\vec{c}_2^2))~. 
\label{gl}
\end{multline}
All collision integrals $\gamma$ and $\gamma_l$ depend on $a_l$ for
all $l$ via $\tilde{\rho}$.
We mention here that our approach is equivalent to the dynamics proposed in
\cite{poeschel}, but has the advantage to give immediately the explicit time
dependence of all coefficients, at least formally.

\subsection*{Collision frequency}
The set of equations (\ref{eq:mom1}) and (\ref{tima})  for $T$ and $a_l$ with
$\gamma_l$ and $\gamma$ given by (\ref{gl}) and (\ref{eq:gg}) are the
main results of this paper.  However, before analyzing them in detail
in the next section and comparing them with computer simulations in
Sec.~5, it is instructive to study the collision frequency $\omega$.
In order to do so, we introduce the average number of collisions,
$\tau$, that a particle has suffered in a time $t$.  Then, the
collision frequency is defined as $\omega=\frac{d}{dt}\tau(t)$.  In
elastic fluids $\omega$ is a constant number depending only on the
density and temperature, so that $\tau$ and $t$ are proportional
quantities.  In granular fluids, however, $\omega$ depends on time, as
the temperature (and more precisely also the shape of the distribution
function) of the system changes with time. Therefore, it is more
natural from a physical point of view to express the time evolution
equations in terms of the variable $\tau$ rather than $t$.  Moreover,
the hydrodynamic matrix become time independent when the hydrodynamic
equations are expressed in the variable $\tau$ \cite{Cahn}.

 To determine $\omega=\frac{d}{dt} \tau(t)$ we use 
Eq.~(\ref{eq:bgl4}) and the fact that in each collision the number of
collisions that each particle has suffered increases by one and we
obtain \cite{chapcowl}
\begin{gather}\label{taudef}
  \frac{d}{dt}  \tau = \omega_0 \gamma_\tau ~,\quad{\rm and} \\
  \gamma_\tau= \frac{\sqrt{2\pi}}{S_d}\frac{1}{\pi^d}\int d
  \vec{c}_1 d \vec{c}_2 d \hat{\vec{\sigma}} \Theta(\vec{c}_{12}\cdot
  \hat{\vec{\sigma}})(\vec{c}_{12}\cdot\hat{\vec{\sigma}})
  \tilde{\rho}(\vec{c}_1) \tilde{\rho}(\vec{c}_2)~.
\end{gather}
$\gamma_\tau$ depends on all $a_l$ and for the case that all
$a_l=0$ for $l>1$  we would get $\gamma_\tau=1$ and thus the Enskog value
$\omega_0$.  We  define a time $\tilde{\tau}$ by 
\begin{equation}
  \label{tildetaudef}
d \tilde{\tau} = \omega_0 dt~.  
\end{equation}
Note that $\tilde{\tau}$ is only an approximation of $\tau$ defined in
Eq.~(\ref{taudef}), so it does not really measure time in collisions,
but we will show later that the deviations of $\tilde{\tau}$ from $\tau$
remain small for not too high inelasticities. In other words we hope
that the collision frequency is approximately determined by the Enskog
value and corrections due to deviations from the Gaussian affecting
the collision frequency are small.

\subsection*{Cooling rate}
How can the dynamics be described in a state where all coefficients
have already reached their asymptotic values?  Note that the
quantities $\gamma$ and $\gamma_\tau$ are entirely given by the values
of $a_l$. Assuming that all $a_l$ have reached their asymptotic values
for some time $t>t^*$ or equivalently $\tau>\tau^*$ the quantities
$\gamma$ and $\gamma_\tau$ also remain constant and we denote their
asymptotic values by $\gamma^*$ and $\gamma_\tau^*$.
Then we consider Eq.~(\ref{eq:mom1}) and (\ref{taudef}):
\begin{gather}
  \frac{d}{d t} T = - 2 \gamma^*  \omega_0(T)  T~, \\
  \frac{d}{dt} \tau = \omega_0(T)\gamma_\tau^*~,
\end{gather}
which is solved analytically 
\begin{multline}
  T = \frac{T(t^*)}{\big[1+\gamma^*\omega_0\big(T(t^*)\big)
  \big(t-t^*\big)\big]^2} = \\ T(\tau^*) \exp\left(-2
  \gamma^*/\gamma_\tau^* (\tau-\tau^*)\right) ~,
\label{Tti} 
\end{multline}
so that
\begin{equation}
  \label{tauti}
  \tau(t)-\tau^* = \frac{\gamma_\tau^*}{\gamma^*} 
  \ln\big[1+\gamma^*\omega_0\big(T(t^*)\big)
  \big(t-t^*\big)\big]~.
\end{equation}
Eq.~(\ref{tauti}) provides a relation between collisional time
and real time. 
Eq.~(\ref{Tti}), i.e. the algebraic decay of the temperature in time
or the exponential behavior in $\tau$ is called Haff's law.
Furthermore, this equation makes explicit the fact that the shape of
the velocity distribution function modifies the energy decay rate.

\section{Analytical Results}

\subsection*{Truncation scheme}

Up to now we have determined the {\em full} time dependence of the HCS
in terms of the time dependence of all its moments in 
Eqs.~(\ref{eq:mom1}) and (\ref{tima}). This infinitely large system of
differential equations can only be solved 
by truncation. The approximate solution found by truncation only makes sense 
if all neglected terms are {\em small} 
as compared with the remaining ones. 
On the other hand, for small inelasticities the velocity distribution
function is close to a Gaussian, so that $a_2 \ll 1=a_0$.
We generalize this unequality and assume that 
$a_{l+1}\ll a_l$ for all $l$,
i.e. contributions from higher order coefficients 
get smaller the higher the index. 
Therefore, we propose a truncation scheme in which we 
assume that $a_l$ is of order $\lambda^l$, where 
$\lambda$ is a small parameter. 
If we now make ``an approximation of order  ${\cal
  O}(\lambda^l)$'' we neglect {\em all} terms in {\em all} considered
equations higher than $\lambda^l$.            
This truncation scheme produces a finite set of differential equations
that can be solved. 

We will concentrate on two aspects: 
(i) To investigate the {\em dynamics} we integrate the full set of {\em
differential} equations (up to a certain order $\lambda$). The {\em
asymptotic} values of the coefficients can then be obtained by taking
the long-time limit if they become stationary in time.  (ii) To
discuss the stationary state we set the left hand side of
Eq.~(\ref{tima}) equal to zero. This set of coupled and, as the case may
be, non-linear equations can be solved with the numerical tool
provided by the computer algebra program.  Note that not all of these
{\em stationary} values are necessarily dynamically stable solutions
of the corresponding differential equation.

\subsection*{Results to order 2}
\label{simex}
In a first step we only take into account $a_2$ to linear order.  Then
the functional form of the equation for $a_2$ Eq.~(\ref{tima}) is given by
\begin{equation}\label{simex1}
  \frac{d}{dt}a_2 = \omega_0(\gamma_2 + 4 \gamma a_2) \rightarrow
  \frac{d}{d\tilde{\tau}}a_2= \gamma_2 + 4 \gamma a_2~.
\end{equation}
Again, the use of $\tilde{\tau}$ simplifies the form of the equations, as it
eliminates the time dependent factor $\omega_0$. 
We recall here that both $\gamma_2$ and $\gamma$ depend on all
$a_l$, but for the approximation treated here, they only 
depend on $a_2$ in a linear manner. Therefore, we can express 
Eq.~(\ref{simex1}) as
\begin{equation}\label{simexx}
\frac{d}{d\tilde{\tau}}a_2 = A+B a_2 + {\cal O}(\lambda^3),
\end{equation} 
where A and B are
constants given by the collision integrals in $\gamma$ and
$\gamma_2$ with the explicit expressions in two dimensions: 
$ A = \frac{1}{8} (e_n^2-1)(2e_n^2-1) $
and $ B = \frac{1}{128}(30 e_n^4-5e_n^2-32 e_n-57) $.\\
{\it (i) Dynamics--} The evolution equation (\ref{simexx}) 
for $a_2$ where time is expressed in collisions per 
particle is linear in $a_2$, so it can be easily integrated 
to give, when $a_2(0)=0$, 
\begin{equation}
  a_2(\tilde{\tau}) = -\frac{A}{B}(1-\exp(B \tilde{\tau})) ~,
\label{eq:a2}
\end{equation}
so that the asymptotic value of $a_2$ is reached exponentially fast on
a  time scale of the order of $\tilde\tau_0:=-B^{-1}$. 
The decay time  $\tilde\tau_0$ ranges between 1.7  and 2.25. 
As an important consequence the asymptotic solution is quickly reached
on a kinetic time scale of few collisions per particle. \\
{\it (ii) Stationary state--} For times larger than $\tilde\tau_0$ 
$a_2$ reaches the stationary value  of $-A/B$ 
which coincides with the values calculated in \cite{noije}.

\subsection*{Results to order 3}
\label{rto3}
\begin{figure}[htbp]
\begin{center}
  \epsfig{file=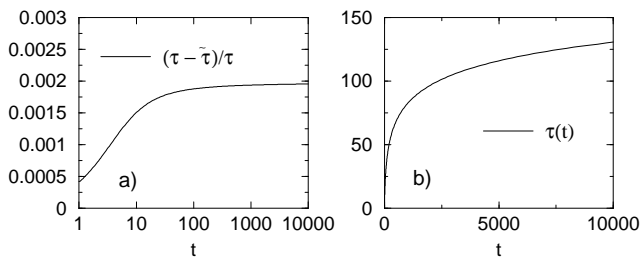,width=0.47\textwidth}
\caption{a) Relative deviations of collisions per particle from the
  approximation given by the Enskog value as a function of time.  b)
  Collisions per particle as a function of time. Dissipation, $e_n=0.9$.  }\label{fig:1}
\end{center}
\end{figure}

To keep the discussion simple and to compare results from order to
order, we first take into account only $a_2$ and $a_3$ i.e.  up to ${\cal
O}(\lambda^3)$. We then still have to deal with equations which are
linear in the coefficients ($a_2^2$ is already of order ${\cal
O}(\lambda^4)$). In the next section we will discuss the non-linear
case up to order ${\cal O}(\lambda^6)$.  The equations read 
\begin{eqnarray} \label{sys3a}
  \frac{d}{dt} T &=& - 2 \gamma \omega_0 T~, \nonumber \\
  \frac{d}{dt} a_2 &=& w_0 \gamma_2 + 4 \gamma \omega_0 (a_2 -0)~, \nonumber \\
  \frac{d}{dt} a_3 &=& w_0 \gamma_3 + 6 \gamma \omega_0 (a_3 -a_2)~, \nonumber \\
  \frac{d}{dt} \tau &=& \omega_0 \gamma_\tau~, \label{sys3}  \\
  \frac{d}{dt}  \tilde{\tau} & = & \omega_0  \quad \text{ neglecting
    corrections of $a_2$ and $a_3$.}\nonumber  
\end{eqnarray}
We use computer algebraic
programs to calculate the collisions integrals $\gamma$, $\gamma_l$
and $\gamma_\tau$ up to order ${\cal O}(\lambda^3)$. 
The analytical solutions are rather lengthy and we will only show
here results  for a system with $e_n =0.9$ in Figs. 1--3.  \\
{\it(i) Dynamics--} We have solved the simultaneous Eqs.\
(\ref{sys3}) numerically\footnote{We have used the built-in
  numerical procedure {\tt dsolve} of MAPLE to integrate the
  differential equation.} for the case $e_n =0.9$ and in the following we
always plot time in units of $1 / \omega_0(T(0))$ and temperature in
units of $T(0)$.  We have chosen $a_2(0)=a_3(0)=0$ as initial
condition.  In a first step we proof that the approximation to use
$\tilde{\tau}$ instead of $\tau$ can be justified (at least to this
order).  In Fig.\  \ref{fig:1} a) we show the relative deviations of
the true number of collisions to the approximation given by the Enskog
Boltzmann value, i.e $ (\tau - \tilde{\tau})/ \tau $ as function of
time in a semilogarithmic plot.

%
%
We see that the relative deviations remain smaller than 0.2 \%.  This
allows us, at least in the homogeneous cooling state, to use
$\tilde{\tau}$ instead of $\tau$ in Eqs.\ (\ref{Tti}) and
(\ref{tauti}).

In the asymptotic state we get $\gamma^*= 0.04723$  
for $e_n=0.9$ and values from the numerical integration of 
Eqs.\ (\ref{sys3}) coincide with Eq.~(\ref{Tti}) and
(\ref{tauti}) within the graphical accuracy, so we plotted here only
the numerical solution.  Fig.\  \ref{fig:1} b) shows $\tau(t)$ which
has the same form as predicted in Eq.~(\ref{tauti}).

\begin{figure}[h]
\begin{center}
  \epsfig{file=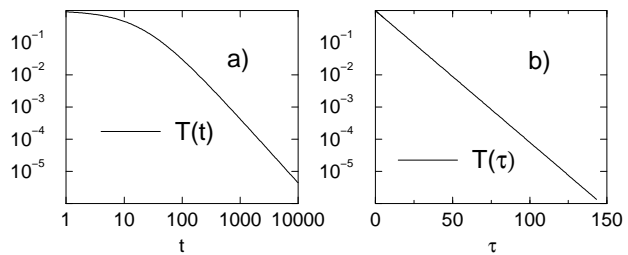,width=0.45\textwidth}
\end{center}
\caption{Temperature as a function of time a) and collisions per
  particle b). Dissipation, $e_n=0.9$.  }\label{fig:2}
\end{figure}

In Fig.\ \ref{fig:2} a), we show $T$ as a function of time in a double
logarithmic plot. We see the well-known asymptotic time dependence
$T\propto t^{-2}$. In Fig.\ \ref{fig:2} b), we show $T$ as a function
of $\tau$ in a semi logarithmic plot resulting in a straight line with
slope $-2\gamma^*$ as predicted by Eq.~(\ref{Tti}).

%
%

\begin{figure}[htbp]
\begin{center}
  \epsfig{file=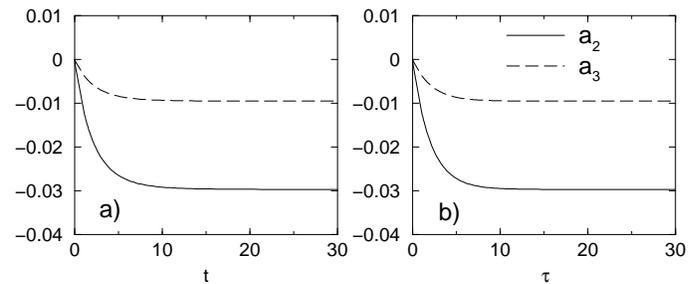,width=0.5\textwidth}
\caption{Coefficients $a_2$ and $a_3$
  as a function of time a) and collisions per particle
  b). Dissipation, $e_n=0.9$. }\label{fig:3}
\end{center}
\end{figure}

In Fig.\ \ref{fig:3} we show the time dependence of $a_2$ and $a_3$
as a function of time a) and as a function of $\tau$ b).

%
%

\begin{figure}[htbp]
\begin{center}
  \epsfig{file=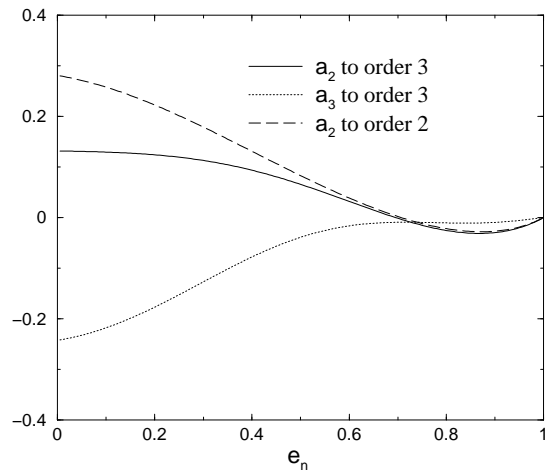,width=.4\textwidth}
\caption{Coefficients $a_2$ to ${\cal O}(\lambda^2)$, and $a_2$ and
$a_3$ to ${\cal O}(\lambda^3)$ as a function of $e_n $}\label{fig:4}
\end{center}
\end{figure}

We see that $a_2$ and $a_3$ reach their asymptotic value on a 
very short time scale which is of the order of few $\tau$'s. 
Therefore, few collisions per particle are necessary
to reach the asymptotic state for $a_2$ and $a_3$.\\
{\it(ii) Stationary state--}
As mentioned above we calculate the stationary values by setting the
l.h.s of Eq.~(\ref{sys3a}) equal to zero. 
In Fig.\  \ref{fig:4} we show the results for the stationary values
of $a_2$ and $a_3$ to ${\cal O}(\lambda^3)$ as well as $a_2$ to ${\cal
  O}(\lambda^2)$.  

%
%

As long as $e_n >0.6$, $a_2$ to ${\cal O}(\lambda^2)$ does not differ
significantly from $a_2$ to ${\cal O}(\lambda^3)$ and $a_3$ remains
small. We see stronger differences for smaller $e_n $ and $a_3$
becomes as important as $a_2$ which indicates stronger deviations from
the Gaussian state. We also cannot assume anymore that corrections of
higher orders remain small since we do not have any indication that the
series is converging in the sense that the $|a_l|$ are small and
decreasing.

Since to ${\cal O}(\lambda^3)$ we have to deal with a set of linear
equations we only find one unique solution.  Considering higher orders
one will find many solutions whose validity must be investigated.  We
will discuss this problem in the next section.

\subsection*{Results to order 6}
\label{rto6}
\begin{figure}[htbp]
  \begin{center} \leavevmode
    \epsfig{file=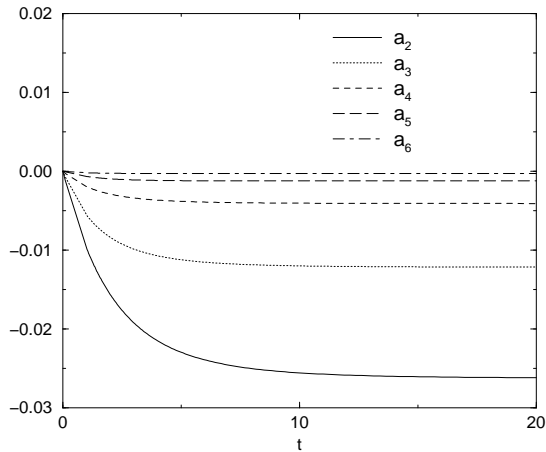, width=0.4\textwidth}
    \caption{The coefficients $a_2,\dots,a_6$ calculated to order
    $\mathcal{O}(\lambda^6)$ as a function of time for $e_n =0.8$.}
    \label{fig:o6} \end{center}
\end{figure}

In this section we go to ${\cal O}(\lambda^6)$, which is the highest
order we were able to calculate with the computer algebra program.\\
{\it (i) Dynamics--} In Fig.\ \ref{fig:o6} we show for $e_n =0.8$ the
dynamics for the 5 non-vanishing coefficients $a_2,\ldots,a_6$ as a
function of time. We have chosen the initial condition
$a_2(0)=\ldots=a_6(0)=0$. We see again a very fast decay to their
asymptotic values. We observe that $|a_l| > |a_{l+1}|$ and in this
sense the perturbation expansion seems to converge.\\ \noindent
%
%

\begin{figure}[htbp]
  \begin{center}
    \leavevmode \epsfig{file=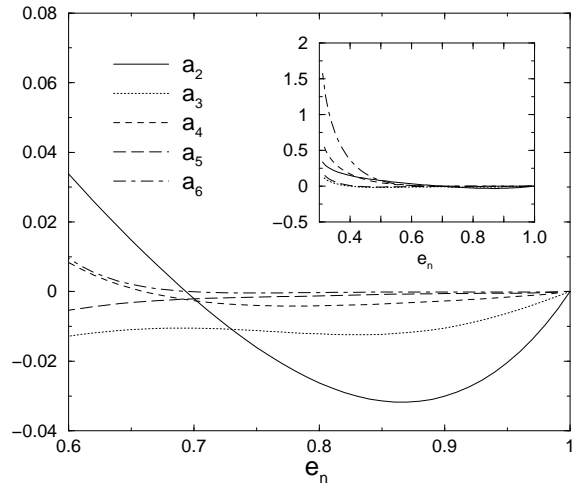, width=0.42\textwidth}
    \caption{Stationary values $a_2,\dots,a_6$ calculated to order
    $\mathcal{O}(\lambda^6)$ as a function of $e_n$}
    \label{fig:o62}
  \end{center}    
\end{figure}

{\it(ii) Stationary state--} We calculate the stationary values
by setting the l.h.s of Eq.~(\ref{tima}) equal to zero.  In Fig.\
\ref{fig:o62} we show the results of the stationary values as a
function of $e_n>0.3$.

%
%

For $e_n>0.7$ the coefficients
remain small and the expansion seems to converge in the sense that
$|a_l| > |a_{l+1}|$ for all $l$. For $e_n<0.7$ the absolute values of
the coefficients start to grow and seem to diverge with $e_n$ approaching
$0.3$. 

\begin{figure}[htbp]
  \begin{center}
    \leavevmode \epsfig{file=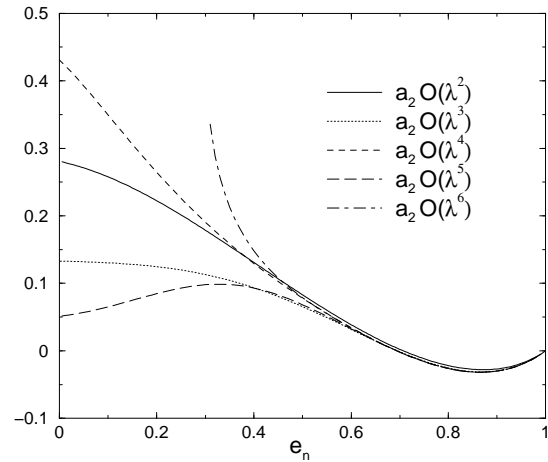,width=0.4\textwidth}
    \caption{  Stationary value of $a_2$ to order
      $\mathcal{O}(\lambda^3)$ up to  $\mathcal{O}(\lambda^6)$ as
      function of $e_n$.}
    \label{fig:o2}
  \end{center}
\end{figure}

To discuss the validity of these results we compare in Fig.\
\ref{fig:o2} the asymptotic values of $a_2$ of order
$\mathcal{O}(\lambda^2)$ up to order $\mathcal{O}(\lambda^6)$. As long
as $e_n>0.6$, we do not find significant differences between the two
orders, only the first order differ slightly from the other ones.
This is a further hint that for these values of $e_n$ the perturbation
method works. Moreover, the ratios of $a_{l+1}/a_{l}$  
are small and of the same order: for instance, for $e_n=0.85$, 
these ratios are: $a_3/a_2=0.39,\, a_4/a_3=0.29,\, a_5/a_4=0.24$ and
$a_6/a_5=0.16$. 

For $e_n<0.6$ the results differ drastically from order
to order and the proposed truncation scheme for Eq.~(\ref{tima}) fails. 
We conjecture that around $e_n \approx 0.6$ an essential 
change in the distribution function
occurs. Then the distribution function is no more described by small
deviations around a Gaussian and might be better expressed by an
expansion around an exponential as suggested in \cite{esipoe}
and confirmed by DSMC  simulations of \cite{brey2}.
We will go back to this point at the end of Sec.~5.

%
%

In view of Fig.\ \ref{fig:o2}, another possible  explanation
for the failure of the convergence is that the expansion in
$\lambda$ is of asymptotic type, in such a way that including  higher 
orders the expansion would break down in the whole range $0<e_n<1$. 
Unfortunately, we cannot decide which is the correct option, as we
can only calculate up to order $\mathcal{O}(\lambda^6)$. 

\subsection*{Further unstable solutions}
Since we have to deal with non-linear equations, the solution is not
unique and e.g. for $e_n =0.8$ two further stationary solutions can be
found, similar as in \cite{poeschel}. We list the values of the other
coefficients for these stationary solutions:
\begin{center}
\begin{tabular}[c]{l|l|l}
  & Solution 1 & Solution 2 \\ \hline $a_2$ & 8.95 & -24.62\\ \hline
  $a_3$ & -14.39 &-4.50 \\ \hline $a_4$ & 59.11 & 39.53\\ \hline $a_5$
  & -109.17 & 178.6\\ \hline $a_6$ & 127.8 & -197.7\\ \hline
\end{tabular}
\end{center}
Both solutions  are   dynamically  unstable which  we have    shown by
numerical  integration  of  the corresponding  differential  equations
(\ref{eq:mom1}) and (\ref{tima}).   In  addition we observe   that the
higher coefficients are not at all negligible so that our assumptions,
which   should   allow us  to  truncate   the  system of  differential
equations,  are severely  violated. Hence for these  cases  we can assume
that we have not even found an approximate solution of the homogeneous
Boltzmann equation.

\section{Computer simulation results}
\label{simr}
In the literature only Direct Simulation Monte Carlo methods (DSMC) have been
used for measuring the values of $a_2$ and $a_3$ \cite{brey} and $a_2$
agrees very well with the value calculated in \cite{noije}.  However,
DSMC lacks some features of the real IHS fluid, as correlations among
the particles.

We present in this paper for the first time results for $a_2$ and for
high-energy tails obtained
from Molecular Dynamics (MD) simulations of the IHS system in 2
dimensions. Our code closely follows the event driven molecular
dynamics code presented in \cite{Allen+Tildesley}, adapted to the
collision rules described in Eqs.~(\ref{coll}) and accelerated by
techniques described in \cite{lub}.  Typical simulations
are performed with $N=50000$ particles in a square box of size $L$,
being its area fraction $\phi=\frac{\pi\sigma^2}{4}n= \frac{\pi
  N\sigma^2}{4L^2}$. The initial configuration is that of an elastic
fluid at equilibrium (Maxwellian distribution for velocities and
equilibrium correlations for the positions due to excluded volume effects),
prepared by running the system with $e_n=1$ (elastic interactions) for
not less than 50 collisions per particle. Therefore, in the initial state
$a_l=0$ for $l\geq 1$.

At the beginning of the inelastic evolution the system remains for
some time in the HCS, where the assumptions made in Sec.~2 are fully
applicable and where computer simulations will serve to test those
predictions. Later, vortices and clusters start to develop through the
system and homogeneity is lost \cite{goldhirsch,noijeprl}.  The
higher the density $\phi$ and the inelasticity the sooner  these
structures appear and important deviations from the theoretical
values of $a_l$ are expected. We will come back to this point later.
Furthermore, the analytical results are {\em independent} of the
density\footnote{To eliminate the dependency of
  real time on the density,  time can be scaled by the collision
  frequency $\omega(T(0))$ at time $t=0$} when expressed in $\tau$,  
but only depend on the
inelasticity $e_n$. Hence we have performed our
simulations at low density of $\phi=0.05$, although simulations at
higher and lower densities have also been
 carried out.

\subsection*{Results for moments}
\begin{figure}[htbp]
\begin{center}
\leavevmode \epsfig{file=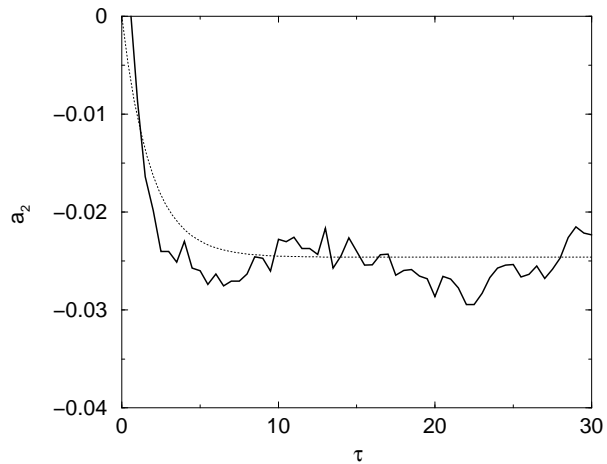,width=0.35\textwidth,angle=270}
\caption{Time evolution of $a_2$ for an IHS system with $\phi=0.05$
and $e_n=0.92$. The dotted line is the analytical result, while the solid
line is the numerical simulation data averaged over two realizations.
}\label{fig-exp1}
\end{center}
\end{figure}

The typical time evolution of the 4th cumulant is shown in
Fig.~\ref{fig-exp1}, where we have plotted the value of $a_2$ versus
the number of collisions per particle $\tau$ for a low density case
$\phi=0.05$ and low inelasticity $e_n=0.92$. The dotted line is the
result of Eq.~(\ref{eq:a2}) while the solid line is the result of the
numerical simulation averaged over two realizations to slightly
improve the accuracy. In this plot we observe the typical features of
the IHS evolution described in former sections. Initially $a_2$ is
equal to zero, as the system starts from a Maxwellian distribution,
with $a_l=0$ for $l>1$. Then, within a very short time of a few
collisions per particle, deviations from a Maxwellian build up in the
system and the asymptotic values of $a_l$ are reached. This is a very
fast process on a hydrodynamic time scale, as it involves only a few
collisions per particle and, therefore, a few mean free times. Then the
moments stay constant (within the accuracy of the computer
simulations) as long as the system remains in the HCS.

%
%

A best fit of the simulation data  to an expression 
like Eq.~(\ref{eq:a2}) gives that
$\tilde\tau_0=1.6 \pm 0.5$ and $a_2(\infty)= -0.026\pm 0.004$, while
the theoretical values developed in Sec.~4  are
$\tilde\tau_0= 1.85$ and $a_2(\infty)=-0.0246$. We observe excellent
agreement with the theory. Unfortunately, the accuracy of our
computer simulations is not high enough to distinguish between the lowest
order and the order 6. The DSMC method also finds good agreement with
the value of $a_2$ \cite{brey}.

%
%
\begin{figure}[htbp]
\begin{center}
\leavevmode \epsfig{file=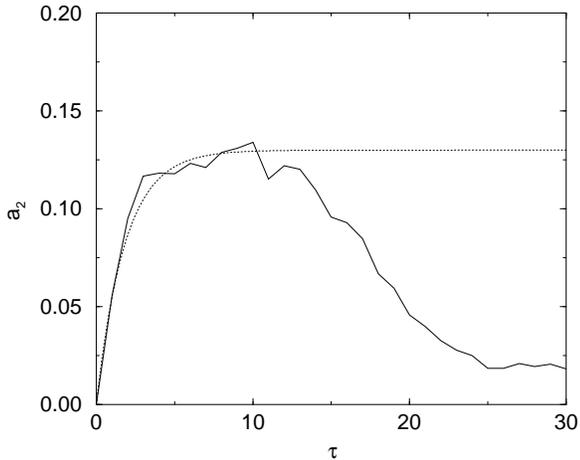,width=0.35\textwidth,angle=270}
\caption{Time evolution of $a_2$ for an IHS system with $\phi=0.03$
and $e_n=0.4$. The dotted line is the analytical result, while the solid
line is the numerical simulation data averaged over two realizations.
Deviations from HCS are seen after $\tau> 12$.
}\label{fig-exp2}
\end{center}
\end{figure}

At later times the system is no longer in the HCS and the assumptions
used in the theoretical sections break down. This is best illustrated
in the Fig.~\ref{fig-exp2}, where a simulation at low density
$\phi=0.03$ but at very high inelasticity $e_n=0.4$ is presented. We
observe the same features described in Fig.~\ref{fig-exp1} with values
from simulations  $\tilde\tau_0=1.5\pm 0.5$ and $a_2(\infty)
=0.125\pm 0.007$, a best fit for $\tau< 10$,  compared to the
theoretical values  $\tilde\tau_0=1.83$ and $a_2(\infty)=0.130$.
Again, the agreement is excellent.
However, after a short time $\tau\mayapr 12$ the values of the moments
start to deviate from the theoretical predictions.  For $\tau \mayapr 12$
the homogeneity assumption breaks down. This can be checked, e.g.  by
plotting the curve of energy vs time \cite{goldhirsch2,mcnamara}:
deviations from Haff's law imply lack of homogeneity \cite{Orza,B+E}.
Visual inspection of the system (not presented here) show that, at
this high inelasticity, the system immediately develops currents and
dense clusters where particles move almost parallel inside them. The
description in terms of $\tilde\rho(c)$ is wrong, as it does not take
into account the local macroscopic currents. The values of $a_2$ can,
at this late evolution stages, grow up to values of $a_2=2$
\cite{goldhirsch2}. This regime is outside the scope of this article.

Concerning the duration of HCS, a hydrodynamic analysis
\cite{Orza} shows that currents develop through the system 
on a time scale of $\tau$ of the order of $ \gamma_0^{-1}$. 
In Fig.~\ref{fig-exp1} the deviation with respect to $a_2(\infty)$
appears  at $\tau\simeq 35= 1.3 \gamma_0^{-1}$, while 
in Fig.~\ref{fig-exp2} it appears at  $\tau\simeq 12= 2.5 \gamma_0^{-1}$.

\begin{figure}[htbp]
\begin{center}
\leavevmode \epsfig{file=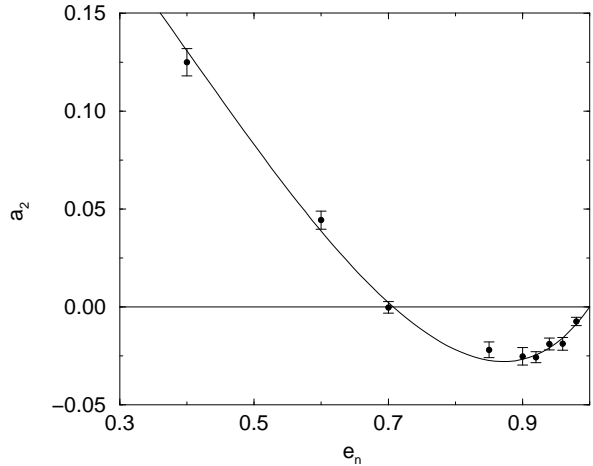,width=0.35\textwidth,angle=270}
\caption{ Coefficient  $a_2$ versus the coefficient of restitution
  $e_n$. The solid
  line is the theoretical prediction of  Eq.~(\ref{simexx}) and the circles
  are the values calculated from MD simulations with their corresponding
  error bars.  }\label{fig-exp3}
\end{center}
\end{figure}

The results of the MD simulations for $a_2$ in the HCS compared with
the asymptotic solution of Eq.~(\ref{simexx}) are given in
Fig.~\ref{fig-exp3}. The agreement is excellent even down to high
inelasticities as $e_n=0.4$, as shown in Fig.~\ref{fig-exp2}. However
our simulations do not have precision enough to test if higher order
corrections are important. 
Concerning the values of $\tilde\tau_0$, MD results are very close,
but always smaller than the theoretical values quoted below
Eq.~(\ref{eq:a2}) and they are affected by large errors.

%
%

To conclude this section, we have also measured the 6th moment,
$\langle c^6\rangle =\langle v^6\rangle /\langle v^2\rangle ^3$, and
from this quantity we have obtained $a_3$ given by  $a_3=
-\frac{1}{6} \langle c^6\rangle + (1+\frac{d}{4}) \langle c^4\rangle 
+(\frac{d^3}{48}-\frac{7d}{12}-1) $. The results
also agree with the solutions of Eq.~(\ref{sys3}), but, as
expected, the discrepancies are now larger because the
absolute value of $a_3$ is very small.

\subsection*{Cooling rate}

Another way to  test the analytical results of Sec.~4 
is to measure the dissipation rate given by the decay of the temperature
$T$ or energy $E$ versus $\tau$. Haff's calculations predicts an exponential
law $\exp(-2\gamma_0 \tau)$ with $\gamma_0=(1-e_n^2)/2d$
while higher order corrections modify it
as shown in Eq.~(\ref{Tti}). These corrections respect to $\gamma_0$  are very
small of only few parts in a thousand. 
If the IHS system is too large or inelastic, these deviations cannot
be measured, as the curve of energy vs $\tau$ bends apart of the exponential
law \cite{mcnamara,B+E}. This is due to the appearance of currents and
vortices as quantitatively explained by \cite{B+E}.  Moreover,
temperature and energy are no longer proportional, and, in contrast to
energy, temperature is difficult to measure in a numerical experiment.

However, if the system is small enough no shear or clustering
instability is excited (see e.g. Refs.\cite{mcnamara,Orza} for
detailed explanations) and the system is forced to remain in the HCS
for all times, where Eq.~(\ref{Tti}) is valid. This is called `kinetic
regime' in Ref.\cite{mcnamara}. The drawback of this method is that,
for a given density, it sets a {\em maximum} number of particles, that
decreases with increasing inelasticity. As described in
Ref.~\cite{Orza} a lower bound for $k_{min}=\frac{2\pi}{L} <
k^*_\perp$ has to be satisfied, in the notation of \cite{Orza}. We
keep $k_{min}=2 k^*_\perp$. This condition, together with our chosen
low density of $\phi=0.05$, restricts the number of particles to
$N=80$ at $e_n=0.70$, and the results are no longer reliable. Even
for $e_n=0.95$ the maximum number of particles is only $N\simeq 300$.
Hence we restrict ourselves to $e_n>0.7$.

%
%
\begin{figure}[htbp]
\begin{center}
\leavevmode \epsfig{file=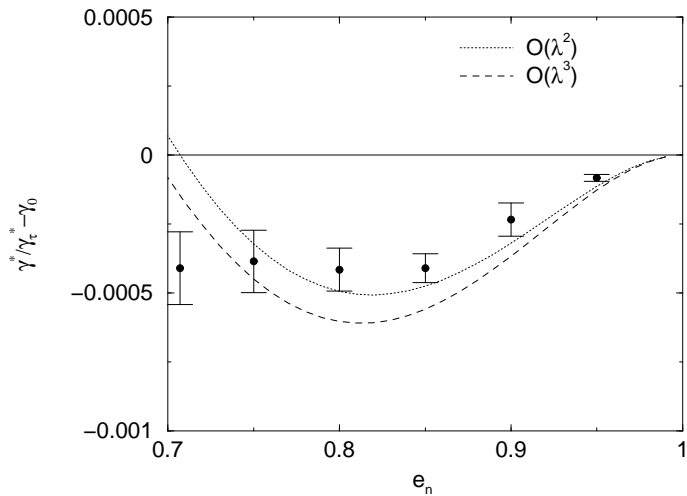,width=0.37\textwidth,angle=270}
\caption{Deviations of the cooling rate as a function of the inelasticity
as obtained in simulations of small systems that remain in the HCS.
}\label{fig-exp4}
\end{center}
\end{figure}

Following this method we have performed simulations and have measured
the energy decay rate as a function of $\tau$. We have
verified that it is indeed exponential for all
times, and,  in order to improve the statistics, results are averaged over
1000 realizations.  The results are presented in Fig. \ref{fig-exp4},
where we plot the difference between the stationary energy decay rate
$\gamma^*/\gamma_\tau^*$ from Eq. (\ref{Tti}) and $\gamma_0$. 
Circles are data obtained by MD simulations with their errors bars, while 
dotted and dashed lines are the results from our theory to order 
${\cal O}(\lambda^2)$ and ${\cal O}(\lambda^3)$ respectively. 
The small number of particles does not allow to obtain
any significant result below $e_n< 0.7$. 
For $e_n > 0.75$ we observe a reasonable agreement, 
although we find significant deviations with respect to the theoretical 
results,  MD data are always smaller that theoretical values. 
Unfortunately, no direct measurements with DSMC  of the deviations respect
to $\gamma_0$ have been reported so far. They would allow us to 
compare our results and elucidate the nature of these deviations.

It is important to note here that due to the small size of the
deviations of the cooling rate with respect to $\gamma_0$ it is
necessary to use in the theory the real number of collisions $\tau$
instead of $\tilde\tau$.  The approximation $\tilde\tau\simeq \tau$ is
correct within a few parts in a thousand as shown in Sec.~4, which is
of the same order of magnitude as the correction
$\gamma^*/\gamma_\tau^*$ with respect to $\gamma_0$. This is not the
case, however, for calculations of the dynamics of $a_l$, for example 
given in Eq. (\ref{eq:a2}), where we are not interested in such small
deviations but want to give a first estimate of the time
scales. Hence  the use of $\tilde\tau$ and the approximation made in Eq.
(\ref{eq:a2}) is fully justified.

Finally, there is still the open question if
deviations from the theoretical $a_2$  and the cooling rate 
are due to the existence of
correlation in the HCS and the breakdown of the  {\em molecular chaos}
hypothesis in Eq.~(\ref{eq:bgl}) reported in \cite{TC,Jago}. These
effects cannot be tested in DSMC either, as this method is based on
the factorization of the two particle distribution function.

\subsection*{High-energy tails}
In Refs. \cite{esipoe,noije} it has been shown that strong deviations with
respect to the Maxwellian are present in the tails of the
distribution, where particles have large energies.
More precisely, they have found if $c=v/v_0(t)\gg (1-e_n^2)^{-1}$ the
velocity distribution function is no longer  Maxwellian, but a simple
exponential $\tilde\rho\simeq {\cal A} \exp(-Ac)$ instead.
This exponential distribution has been verified by numerical solutions
of the Boltzmann equation using the DSMC method \cite{brey2}. The
shape of the exponential is very different to the Maxwellian and
therefore it is understandable that an expansion of the type 
given in Eq.~(\ref{rhotilde}) might be  non-convergent as suggested by our
theoretical analysis. 
Another reasonable possibility is that the series is indeed convergent
but we have not gone high enough in the truncation scheme or there is
a better choice of truncation, because
$a_{l+1}$ is as large as $a_l$, as shown in Fig.~6. 

In order to investigate the velocity distribution function and make
the exponential range accessible, we have performed MD simulations with
extreme inelasticities of $e_n=0.1, 0.2, 0.4$ to compare to moderate
inelasticities $0.6$ and $0.8$. Moreover, as we need high accuracy in
the tails, where populations are small, we have simulated systems with
250~000 particles at $\phi=0.05$. However, even with this large number of particles
we are not able to obtain the accuracy that can be achieved by the
DSMC method \cite{brey2}. We will only be able to give evidences of
the exponential tail.  We measured the distribution function at times,
where $a_2$ has already reached its asymptotic value, but the
homogeneity assumption is still valid. In the example of Fig.
\ref{fig-exp2} this would be at times $5\menapr\tau\menapr 10$.

To estimate the distribution function from data, we use the kernel
estimator technique described e.~g. in \cite{eubank}. In general,
given a set of outcomes $x_i$,\, $i=1,\ldots,M$ of a random experiment the
distribution function $\rho(x)$ can be estimated by
\begin{equation} \label{kernel}
 {\rho}(x) \simeq \frac{1}{M}\sum_{i=1}^{M}
  \frac{1}{\sqrt{2\pi}\delta}\exp\left(\frac{(x_i-x)^2}{2\delta^2}\right)~.
\end{equation}
The idea behind this method is that each data point also gives some
information about its surrounding, which can be justified for smooth
distribution functions. It is not necessary to use a Gaussian as
kernel in Eq.~(\ref{kernel}), any normalized and more or less sharply
peaked function can be used. The value $\delta$ is a free parameter
which was chosen such that the measured distribution function for the
elastic gas (i.e.  the initial condition) is fitted best (to the eye)
to the Maxwellian.  We choose $\delta=0.05$. Since the distribution
function $\tilde{\rho}(c^2)$ --with $c$ the modulo of the velocity-- is
not continuous at $c=0$, i.e. $\tilde{\rho}=0$ for $c<0$ and
$\tilde{\rho}(0) \approx 1/\pi $ this technique gives bad results
around $c=0$, but better results than the histogram method for the
interesting high-velocity limit, where only few data points are given.

\begin{figure}[htbp]
  \begin{center}
    \leavevmode
\epsfig{file=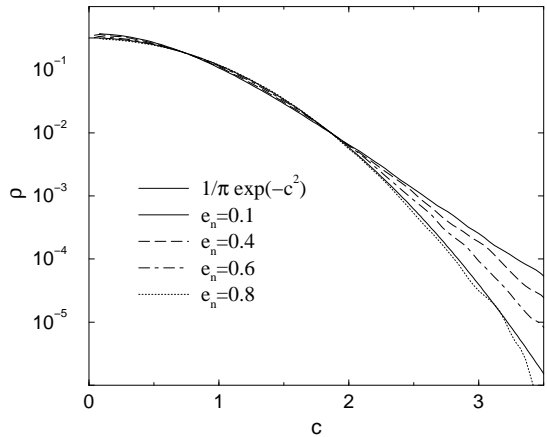, width=0.4\textwidth}
\caption{Maxwellian and measured velocity distribution function
  $\tilde{\rho}$ as a function of $c$ for various values of
  $e_n$.}\label{histo}
  \end{center}
\end{figure}

The simulation results for the velocity distribution function
$\tilde{\rho}$ as a function of $c$ as well as the Maxwellian are
shown in Fig. \ref{histo} in a semi-logarithmic plot. For $c\mayapr
3.5$ the statistical accuracy is poor, so results are only plotted up
to $c=3.5$.

%
%

We observe that the deviations with respect to the Maxwellian are
larger for lower values of $e_n$. On the contrary, as $e_n$ increases, the
measured distribution approaches the Maxwellian.  Closer inspection
shows that for $e_n=0.1$ the distribution gets possibly close to an
exponential (straight line in the semilogarithmic plot of Fig.
\ref{histo}) for $2\menapr c \menapr  3.5$, while for $e_n=0.6$ the
range where $\log\tilde{\rho}$ seems to be linear shrinks to
$3\menapr c\menapr  3.5$. We will show below that in this case
($e_n=0.6$ and $c \menapr  3.5$) the distribution function can be
reasonably well described by the results of the perturbation expansion
around the Gaussian given in Eq. (\ref{rhotilde}).  If we perform a
linear fit to an exponential in these ranges, we obtain values of the
coefficient $A$ quite close to those reported by \cite{noije}, tested
in \cite{brey2} by DSMC method.  For instance, for $e_n=0.1$, $A\simeq
-3.2$, that increases to $A\simeq -3.8$ at $e_n=0.4$ 
and further to $A\simeq -4.7$ at $e_n=0.6$.

\begin{figure}[htbp]
\begin{center}
    \leavevmode
\epsfig{file=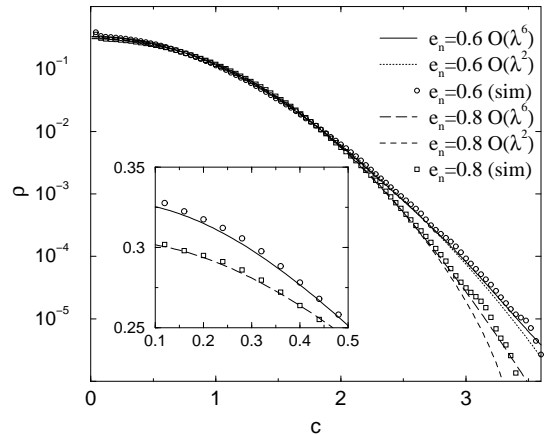, width=0.4\textwidth}
\caption{Measured velocity distribution function
  $\tilde{\rho}$ as a function of $c$ for $e_n=0.6$ and $e_n=0.8$
  compared to results of Sec.~4 to orders $\lambda^2$ and $\lambda^6$. The
  inset shows an blow up of the results for small velocities. }
\label{histo3}
  \end{center}
\end{figure}

If we go beyond $e_n>0.6$ the perturbation expansions of 
Sec.~4 seems to converge and it make sense to compare the
analytical results with the simulation data. In Fig.~\ref{histo3} we
show for $e_n=0.6$ and $e_n=0.8$ results of the simulations and of the
analytical theory to orders $\lambda^2$ and  $\lambda^6$.

%
%

In the whole range of $c$ which is plotted, the analytical results
to order $\lambda^6$ coincides fairly well with the measured data. 
However, results to order $\lambda^2$ agree at small velocity,
$c \menapr  3$ for  $e_n=0.8$, but fails for higher velocities.
It seems reasonable that higher orders are needed to describe higher 
energy tails, where deviations respect the Gaussian are larger. 

Therefore, for $e_n>0.6$ there is no need to
describe data by an exponential tail for high velocities, although it
cannot be assured, that the theory does not fail for even higher
velocities. Note that the distribution function at $c \approx 3.5$ is
already smaller than $10^{-5}$ so that for 250~000 particles only  2
or 3 particles might not be correctly  described.

\section{Conclusion}
\label{conc}

In this article we investigated by means of analytical theory and
simulations the {\rm dynamics} of a freely cooling system of smooth
granular particles as long as it remains homogeneous. Starting from a
pure Gaussian state the system develops on a fast time scale to a
state where the deviations from the Gaussian (described by cumulants)
are stationary in time and the dynamics is entirely described by a
decreasing kinetic energy.

More technically, we determined formally the {\em full} dynamics of the
homogeneous cooling state in terms of the dynamics of the temperature
$T(t)$ and the {\em time dependent} coefficients $a_l(t)$ of an
expansion of the velocity distribution function in Generalized
or Associated Laguerre polynomials around the Gaussian state. We obtained an
infinitely large system of non linear ordinary differential equations,
which can be solved numerically under the assumption that higher
coefficients do not contribute.

Analytically, we found two main results. i) As far as dynamics is
concerned, the HCS is characterized by the fact that only a few collisions
per
particle are necessary to reach a state where the coefficients are
stationary in time. Then the entire time dependence is given by a slow
algebraically decay of the temperature obeying $\frac{d}{dt} T = - 2 
\omega_0 \gamma T $, with $\gamma$ depending on all the asymptotic values
of the coefficients $a_l$. ii) As far as the asymptotic values of the
coefficients are concerned, the expansion seem to converge in the
sense that $|a_{l+1}|<|a_l| $ for ${e_n}>0.6$ and for this range of
$e_n$ we do not find significant differences between orders.  There
exist further stationary but dynamically instable solutions of the
considered differential equation, which are far away from the
assumption of absolutely decreasing and small coefficients, so 
we cannot make any prediction for that cases. 
For ${e_n}<0.6$ the perturbation procedure seems to fail, the
assumption we made to truncate the system of differential equations
are severely violated and we find a strong dependency on the order of
approximation. We have no answer if going to higher order or choosing
a more suitable truncation scheme would show that the perturbation 
procedure nevertheless works, or, on the other hand, if the expansion 
presented here is only of an asymptotic type. 
A reasonable conclusion is that the system develops to a
state which is very far from a Gaussian and might be better described
by an expansion around an exponential as discussed in \cite{poeschel}
and \cite{brey}.

Although much numerical work has been done on the HCS and clustering 
regimes (see, e.g. \cite{goldhirsch}, \cite{mcnamara},
\cite{Orza}, \cite{TC} and \cite{Stefan}), for the first time event-driven 
simulations are used in the
present work to investigate deviations from the Gaussian distribution 
in the HCS.  Mainly, three aspects were considered: 1) The
dynamics and asymptotic values of the coefficients, 2) the influence
on the decay rate of the temperature, 3) the shape of the velocity
distribution function.
\begin{enumerate}
\item As long as the system remains in the homogeneous state the
  dynamical behavior as well as the static value of $a_2$  
  could be confirmed. Nevertheless, the statistics was too poor to 
  distinguish if higher order analytical results give better values. 
\item Similarly, the decay of the temperature was measured showing
  deviations with respect to the analytical results and opening the 
  possibility to study other effects as correlations. Here 
  we had to assure that the system
  remains in the homogeneous state, restricting the number of
  particles and therefore the quality of the statistics.
\item Measuring the full velocity distribution function we found that
  the  distribution function can be described very well by the
  expansion around the Gaussian as long as $e_n>0.6$. For smaller
  $e_n$ the high-energy tails show an exponential shape confirming
  previous results found by analytical theory \cite{esipoe,noije} and
  DSMC simulations \cite{brey2}.
\end{enumerate}

A possible extension of our work are systems of rough spheres with
constant coefficient of restitution or Coulomb friction. Strong
deviations from the Gaussian are observed in the angular velocity
distribution function, surprisingly for the cases where the particles
are almost {\em smooth} \cite{huthmann,herbst}. It would be
interesting to perform a similar dynamical analysis along these lines.


\begin{thebibliography}{99}

\bibitem{goldhirsch} I. Goldhirsch and G. Zanetti, Phys. Rev. Lett. 
{\bf 70} (1993) 1619. 
\bibitem{goldhirsch2} I.~Goldhirsch, M.-L.~Tan  and G.~Zanetti,
J.~Sci.~Comput. {\bf 8} (1993)  1. 
\bibitem{deltour} P. Deltour and J.L. Barrat, J.~Phys.~I France {\bf 7}
(1997) 137.
\bibitem{esipoe}
S.E. Esipov and T. P\"oschel, J. Stat. Phys. {\bf 86} (1997)  1385. 
\bibitem{mcnamara} S. McNamara and W.R. Young, Phys. Rev. E {\bf 53} (1996)
5089. 
\bibitem{noijeprl}  T.P.C. van Noije, M.H. Ernst, R. Brito and 
J.A.G. Orza, Phys. Rev. Lett. {\bf 79}  (1997)  411.
\bibitem{Orza} J.A.G.~Orza, R.~Brito, T.P.C.~van Noije and M.H. Ernst, 
{ Int.~J.~Mod.~Phys.~C} {\bf 8} (1997) 953. 
\bibitem{Cahn} T.P.C.~van Noije, M.H.~Ernst and R.~Brito,
Phys.~Rev.~E {\bf 57} (1998)  R4891.
T.P.C. van Noije and M.H. Ernst, Phys.~Rev.~E,
in press, Feb. 2000, cond-matt/9907012.
\bibitem{goldshtein} A. Goldshtein and M. Shapiro, J. Fluid Mech. {\bf 282}
(1995) 75.  
\bibitem{noije}
T.P.C. van Noije and M.H. Ernst, Gran.~Matt. {\bf 1} (1998) 57.
\bibitem{poeschel}
N.~V. Brilliantov and T. P\"oschel, Phys. Rev. E {\bf 61} (2000) 2809. 
\bibitem{brey} J.J.~Brey, M.J.~Ruiz-Montero and D.~Cubero, 
Phys. Rev. E {\bf 54} (1996) 3664.
\bibitem{brilliantov} N.V. Brilliantov and T. P\"oschel, cond-mat/9911212 (1999).
\bibitem{brey2} J.J.~Brey, D.~Cubero and M.J.~Ruiz-Montero, 
Phys. Rev. E {\bf 59} (1999) 1256.
\bibitem{losert} W. Losert, D.G.W. Cooper, J. Deltour, A. Kudrolli
and J.P. Gollub, cond-mat/9901203.
\bibitem{Andres} J.M.~Montanero and A. Santos, Granular Matter, to appear 
(cond-mat/0002323).
\bibitem{oberhettinger}
W. Magnus, F. Oberhettinger, and R.~P. Soni, {\em Formulas and
  Theorems for the Special Functions of Mathematical Physics}, Springer (1966).
\bibitem{chapcowl}
S. Chapman and T.~G. Cowling, {\em The Mathematical Theory of Nonuniform
Gases}, Cambridge University Press, London (1960).
\bibitem{haff} P.K.~Haff, J.~Fluid Mech. {\bf 134} (1983) 401.
\bibitem{Allen+Tildesley} M.P.~Allen and D.J.~Tildesley, 
{\em Computer Simulation of Liquids}, Clarendon Press, Oxford (1989).
\bibitem{lub} B. D. Lubachevsky, J. of Comp. Phys. {\bf 94} (1991) 255.
\bibitem{B+E}  R.~Brito and M.H.~Ernst, 
{ Europhys.~Lett.} {\bf 43} (1998)  497.
\bibitem{TC} S. Luding, M. M\"uller and  S. McNamara, in {\em World Congress on Particle
Technology} (Inst. of Chem. Eng., Davis Building, 165-189 Railway Terrace,
Rugby CV21 3HQ, UK) (1998), ISBN 0-85295-401-9.
\bibitem{Jago} J.A.G.~Orza and R.~Brito, in preparation.
\bibitem{eubank} S.~G. Eubank and J.~D. Farmer, in {\em Introduction
    to nonlinear physics} p. 106-151,  (Springer,  New York,  1997). 
\bibitem{Stefan} S.~Luding, M. Huthmann, S. McNamara and A. Zippelius,
Phys. Rev. E  {\bf 58} (1998) 3416.
\bibitem{huthmann} T.~Aspelmeier, M.~Huthmann  and  A.~Zippelius,
to be published in {\em Granular Matter},
Lecture Notes in Physics, eds. S. Luding and T. P{\"o}schel,
Springer (2000).
\bibitem{herbst} 
 O.~Herbst, M.~Huthmann, and A.~Zippelius, 
cond-mat/9911306 (1999).
\end{thebibliography}
\end{document}